%% file: susy07HA.tex
\documentclass[epj]{svjour}

\usepackage{graphicx}
\usepackage{fancyhdr}
\def\makeheadbox{{
 }}
\input{paperdef}

\def\mytitle{My title} 
\def\myauthors{My name}  
\def\mytype{My type of session}
\def\mysession{My session}

\def\mytitle{Heavy MSSM Higgs Bosons at CMS \ldots} 
\def\myauthors{S.~Heinemeyer, A.~Nikitenko, G.~Weiglein}   
\def\mytype{Contributed Talk}    
\def\mysession{Colliders - Higgs Phenomenology}

\begin{document}

\input susy07HA_main

\end{document}

%% file: susy07HA_main.tex
\graphicspath{{figs/}}

\title{Heavy MSSM Higgs Bosons at CMS:\\ 
``LHC wedge'' and Higgs-Mass Precision}
\author{S. Heinemeyer\inst{1}
\thanks{\emph{Email:} Sven.Heinemeyer@cern.ch}%
 \and
 A.~Nikitenko\inst{2}
 \and
 G.~Weiglein\inst{3}
\thanks{\emph{Email:} Georg.Weiglein@durham.ac.uk}%
}                     
\institute{Instituto de Fisica de Cantabria (CSIC-UC), Santander, Spain
\and Imperial College, London, UK; on leave from ITEP, Moscow, Russia
\and IPPP, University of Durham, Durham DH1~3LE, UK
}
%
\date{}
\abstract{
\input susy07HA_abstract

\PACS{
      {14.80.Cp}{Non-standard-model Higgs bosons}   \and
      {12.60.Jv}{Supersymmetric models}
     } 
} 
\maketitle
%


\section{Introduction}

Identifying the mechanism of electroweak symmetry
breaking will be one of the main goals of the LHC. 
The most popular models are the Higgs mechanism within the Standard
Model (SM) and within the Minimal Supersymmetric Standard Model
(MSSM)~\cite{mssm}. Contrary to the case of the SM, in the MSSM 
two Higgs doublets are required.
This results in five physical Higgs bosons instead of the single Higgs
boson of the SM. These are the light and heavy $\cp$-even Higgs bosons, $h$
and $H$, the $\cp$-odd Higgs boson, $A$, and the charged Higgs boson,
$H^\pm$.
The Higgs sector of the MSSM can be specified at lowest
order in terms of the gauge couplings, the ratio of the two Higgs vacuum
expectation values, $\tb \equiv v_2/v_1$, and the mass of the $\cp$-odd
Higgs boson, $\MA$.
Consequently, the masses of the $\cp$-even neutral Higgs bosons and the
charged Higgs boson are dependent quantities that can be
predicted in terms of the Higgs-sector parameters. Higgs-phenomenology
in the MSSM is strongly affected by higher-order corrections, in
particular from the sector of the third generation quarks and squarks,
so that the dependencies on various other MSSM parameters can be
important.

The current
exclusion bounds within the MSSM~\cite{LEPHiggsMSSM,D0bounds,CDFbounds}
and the prospective sensitivities at
the LHC are usually displayed in terms of the parameters $\MA$ and $\tb$
that characterize the MSSM Higgs sector at lowest order. The other MSSM
parameters are conventionally fixed according to certain benchmark
scenarios~\cite{benchmark2,benchmark3}. We focus here~\cite{cmsHiggs} on
the $5\,\si$ discovery contours for heavy MSSM Higgs bosons, i.e.\ the
lower bound of the ``LHC wedge'', within the ``$\mhmax$~scenario''. 
For the interpretation of the exclusion bounds and prospective discovery
contours in the benchmark scenarios it is important to assess how
sensitively the results depend on those parameters that have been fixed
according to the benchmark prescriptions.
Consequently, we investigate how the 
5$\,\si$ discovery regions in the $\MA$--$\tb$ plane 
for the heavy neutral MSSM Higgs bosons
obtainable with the CMS experiment at the LHC depend on the other MSSM
parameters.


\section{The analysis}

The search for the heavy neutral MSSM Higgs bosons at the LHC will mainly be
pursued in the $b$~quark associated production with a subsequent decay to
$\tau$~leptons~\cite{lhctdrsA,atlashiggs,lhctdrsS}. In the region of
large $\tb$ this production process benefits from an enhancement factor 
of $\TQb$ compared to the SM case. 
The main search channels are%
\footnote{In our analysis we do not consider diffractive
Higgs production, $pp \to p \oplus H \oplus p$~\cite{diffHSM}.
For a detailed discussion of the search reach for the heavy neutral MSSM 
Higgs bosons in diffractive Higgs production we refer to
\citere{diffHMSSM}.}
(here and in the
following $\phi$ denotes the two heavy neutral MSSM Higgs bosons, 
$\phi = H, A$):
\BEA
\label{jj}
    && b \bar b \phi,  \phi \to \tau^+\tau^- \to 2 \, \mbox{jets}\\[.3em]
\label{mj}
    && b \bar b \phi,  \phi \to \tau^+\tau^- \to \mu + \, \mbox{jet}\\[.3em]
\label{ej}
    && b \bar b \phi,  \phi \to \tau^+\tau^- \to e + \,\mbox{jet}
\EEA

The analyses were performed with full CMS detector 
simulation and reconstruction for the following
three final states of di-$\tau$-lepton decays:  
$\tau^+\tau^- \to \,\mbox{jets}$~\cite{CMSPTDRjj},  
$\tau^+\tau^- \to e + \,\mbox{jet}$~\cite{CMSPTDRej}
and $\tau^+\tau^- \to \mu + \mbox{jet}$~\cite{CMSPTDRmj}.
%
The Higgs-boson production in 
association with $b$ quarks, $pp \to b\bar b  \phi$, has been selected using
single $b$-jet tagging in the experimental analysis. The kinematics of the 
$gg \to b\bar b \phi$ production process (2 $\to$ 3) was generated with 
PYTHIA~\cite{PYTHIA}.
The backgrounds considered in the analysis were QCD muli-jet events 
(for the $\tau \tau \to \,\mbox{jets}$ mode), 
$t \bar t, b \bar b$,  Drell-Yan production of $Z, \ga^{\ast}$, $W$+jet, $Wt$
and $\tau \tau b \bar b$. All background processes  
were generated using PYTHIA, except for $\tau^{+} \tau^{-} b \bar b$,
which was generated using CompHEP \cite{Boos:2004kh}.

\begin{table}[htb!]
\renewcommand{\arraystretch}{1.2}
\BC
\begin{tabular}{|c||c|c|c|} \hline
\multicolumn{4}{|c|}
  {$\phi \to \tau^+\tau^- \to \,\mbox{jets}$, 60~\ifb} \\ \hline\hline
$\MA$ [GeV]     & 200   & 500   & 800 \\ \hline
$N_S$           &  63   &  35   &  17 \\ \hline
$\eps_{\rm exp}$ & $2.5 \times 10^{-4}$ & $2.4 \times 10^{-3}$ & 
                                        $3.6 \times 10^{-3}$ \\ \hline
$R_{M_\phi}$     & 0.176 & 0.171 & 0.187 \\ \hline
$\De M_{\phi} / M_{\phi}$ [\%] & 2.2 & 2.8 & 4.5 \\
\hline
\end{tabular}
\EC
\vspace{-0.5em}
\caption{
Required number of signal events, $N_S$, with $\cL = 60$~\ifb\ for a 5$\,\si$
discovery in the channel $\phi \to \tau^+\tau^- \to \,\mbox{jets}$. 
Furthermore given are the
total experimental selection efficiency, $\eps_{\rm exp}$, the ratio of
the di-$\tau$ mass resolution to the Higgs-boson mass, $R_{M_\phi}$, 
and the expected precision of the Higgs-boson mass measurement,
$\De M_{\phi} / M_{\phi}$, obtainable from $N_S$ signal events.
}
\label{tab:jj}
\renewcommand{\arraystretch}{1.0}
\end{table}

\begin{table}[htb!]
\vspace{-3em}
\renewcommand{\arraystretch}{1.2}
\BC
\begin{tabular}{|c||c|c|c|} \hline
\multicolumn{4}{|c|}
  {$\phi \to \tau^+\tau^- \to e + \,\mbox{jet}$, 30~\ifb} \\ \hline\hline
$\MA$ [GeV]      & 200   & 300   & 500   \\ \hline
$N_S$            &  72.9 &  45.5 &  32.8 \\ \hline
$\eps_{\rm exp}$  & $3.0 \times 10^{-3}$ & $6.4 \times 10^{-3}$ & 
                                           $1.0 \times 10^{-2}$ \\ \hline
$R_{M_\phi}$      & 0.216 & 0.214 & 0.230 \\ \hline
$\De M_{\phi} / M_{\phi}$ [\%] & 2.5 & 3.2 & 4.0 \\ \hline
\end{tabular}
\EC
\vspace{-0.5em}
\caption{
Required number of signal events, $N_S$, with $\cL = 30$~\ifb\ for 
a 5$\,\si$ discovery in the channel
$\phi \to \tau^+\tau^- \to e + \,\mbox{jet}$. 
The other quantities are defined as in \refta{tab:jj}.
}
\label{tab:ej}
\renewcommand{\arraystretch}{1.0}
\end{table}

\begin{table}[htb!]
\vspace{-3em}
\renewcommand{\arraystretch}{1.2}
\BC
\begin{tabular}{|c||c|c|} \hline
\multicolumn{3}{|c|}
  {$\phi \to \tau^+\tau^- \to \mu + \,\mbox{jet}$, 30~\ifb} \\ \hline\hline
$\MA$ [GeV]     & 200  & 500  \\ \hline
$N_S$           &  79  &  57  \\ \hline
$\eps_{\rm exp}$ & $7.0 \times 10^{-3}$ & $2.0 \times 10^{-2}$ \\ \hline
$R_{M_\phi}$     & 0.210 & 0.200 \\ \hline
$\De M_{\phi} / M_{\phi}$ [\%] & 2.4 & 2.6 \\ \hline
\end{tabular}
\EC
\vspace{-0.5em}
\caption{
Required number of signal events, $N_S$, with $\cL = 30$~\ifb\ for a 5$\,\si$
discovery in the channel $\phi \to \tau^+\tau^- \to \mu + \,\mbox{jet}$.
The other quantities are defined as in \refta{tab:jj}.
}
\label{tab:mj}
\renewcommand{\arraystretch}{1.0}
\vspace{-1em}
\end{table}

The results 
quoted in \reftas{tab:jj} -- \ref{tab:mj} for the required number of
signal events depend only on the Higgs-boson
mass, i.e.\ the event kinematics,
but are independent of any specific MSSM scenario. 
In order to determine the 5$\,\si$ discovery contours in the
$\MA$--$\tb$~plane these results have to be confronted with the MSSM
predictions.
The number of signal events, $N_{\rm ev}$, for a given parameter
point is evaluated via
\BE
N_{\rm ev} = \cL \times \si_{b\bar b\phi} \times 
             \br(\phi \to \tau^+\tau^-) \times \br_{\tau\tau} \times
             \eps_{\rm exp}~.
\EE
Here $\cL$ denotes the luminosity collected with the CMS detector, 
$\si_{b\bar b\phi}$ is the Higgs-boson production cross section, 
$\br(\phi \to \tau^+\tau^-)$ is the branching ratio of
the Higgs boson to $\tau$~leptons, 
$\br_{\tau\tau}$ is the product of the branching ratios of the two
$\tau$~leptons into their respective final state, 
\BEA
\br(\tau \to \,\mbox{jet} + X) &\approx& 0.65~, \\
\br(\tau \to \mu + X) \approx \br(\tau \to e + X) &\approx& 0.175~,
\EEA
and $\eps_{\rm exp}$ denotes the total experimental selection efficiency
for the respective
process (as given in \reftas{tab:jj} -- \ref{tab:mj}).
For our numerical predictions of total cross sections (see
\citere{sigmaFH} and references therein) and
branching rations of the MSSM Higgs bosons we use the program
{\tt FeynHiggs}~\cite{feynhiggs,mhiggslong,mhiggsAEC,mhcMSSMlong}.
We take into account effects from higher-order corrections and 
from decays of the heavy Higgs bosons into supersymmetric particles.

In spite of the escaping neutrinos, the Higgs-boson mass can be 
reconstructed in the $H,A \to \tau \tau$ channel from the visible $\tau$ 
momenta ($\tau$ jets) and the missing transverse energy, 
$E_{\rm T}^{\rm miss}$, using the collinearity approximation for neutrinos from
highly boosted $\tau$'s. 
In the investigated region of $\MA$ and $\tb$ the two states $A$ and $H$ are  
nearly mass-degenerate.
For most values of the other MSSM parameters the mass difference of $A$
and $H$ is much smaller than the achievable mass resolution, and the 
difference in reconstructing the $A$ or the $H$ will have no
relevant effect on the achievable accuracy in the mass determination.
The precision $\De M_\phi/M_\phi$ shown in \reftas{tab:jj} -- \ref{tab:mj}
is derived for the border of the parameter space in
which a 5$\,\si$ discovery can be claimed, i.e.\ with $N_S$ observed Higgs
events. The statistical accuracy of the mass measurement has been evaluated
via 
\BE
\label{eq:precM}
{\De M_\phi}/{M_\phi} = {R_{M_\phi}}/{\sqrt{N_S}}~.
\EE
A higher precision can be achieved if more than $N_S$ events are observed. The
corresponding estimate for the precision is obtained 
by replacing $N_S$ in \refeq{eq:precM} by the number of observed
signal events, $N_{\rm ev}$. It should be noted that the prospective 
accuracy obtained from \refeq{eq:precM} does not take into account 
the uncertainties of the jet and missing $E_{\rm T}$ energy scales. In 
the $\tau^+\tau^- \to \,\mbox{jets}$ mode these effects
can lead to an additional 3\% uncertainty in the mass 
measurement~\cite{CMSPTDRjj}.


\section{Numerical results for the LHC wedge}

We have evaluated $N_{\rm ev}$ in the $\mhmax$ benchmark
scenario~\cite{benchmark2,benchmark3} as a function of $\MA$ and $\tb$.
For fixed $\MA$ we have varied $\tb$ such that $N_{\rm ev} = N_S$ (as given in
\reftas{tab:jj} -- \ref{tab:mj}). This $\tb$ value is then identified as 
the point on the 5$\,\si$ discovery contour corresponding to the chosen
value of $\MA$.
In this way we have determined the 5$\,\si$ discovery contours
for the $\mhmax$ scenario for $\mu = \pm 200, \pm 1000 \gev$.
\footnote{
A corresponding analysis in benchmark scenarios fulfilling 
cold dark matter constraints can be found in \citere{ehhow}.
}%

\begin{figure}[htb!]
\BC
\includegraphics[width=.4\textwidth,height=6.1cm]{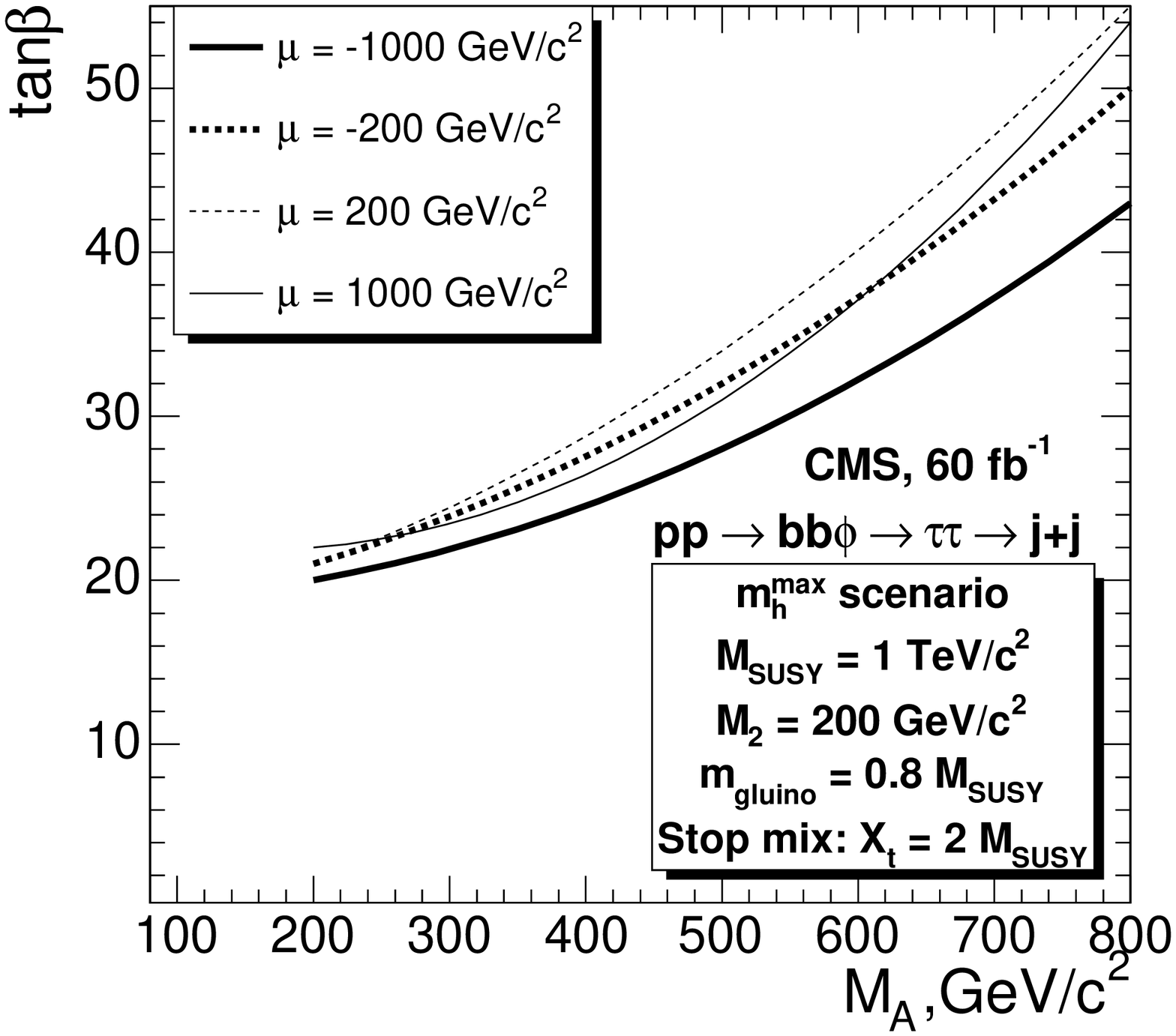}
\includegraphics[width=.4\textwidth,height=6.1cm]{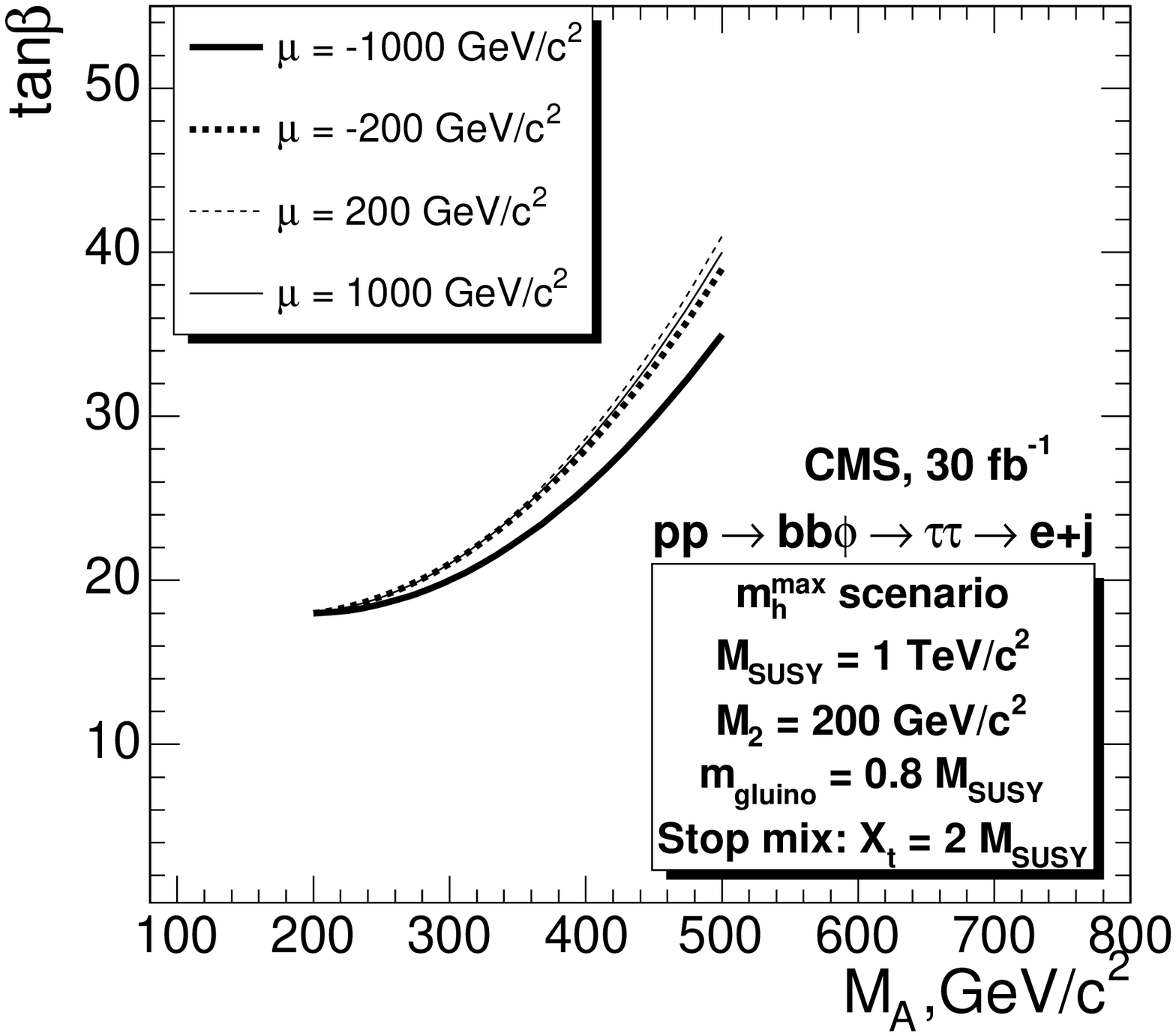}
\includegraphics[width=.4\textwidth,height=6.1cm]{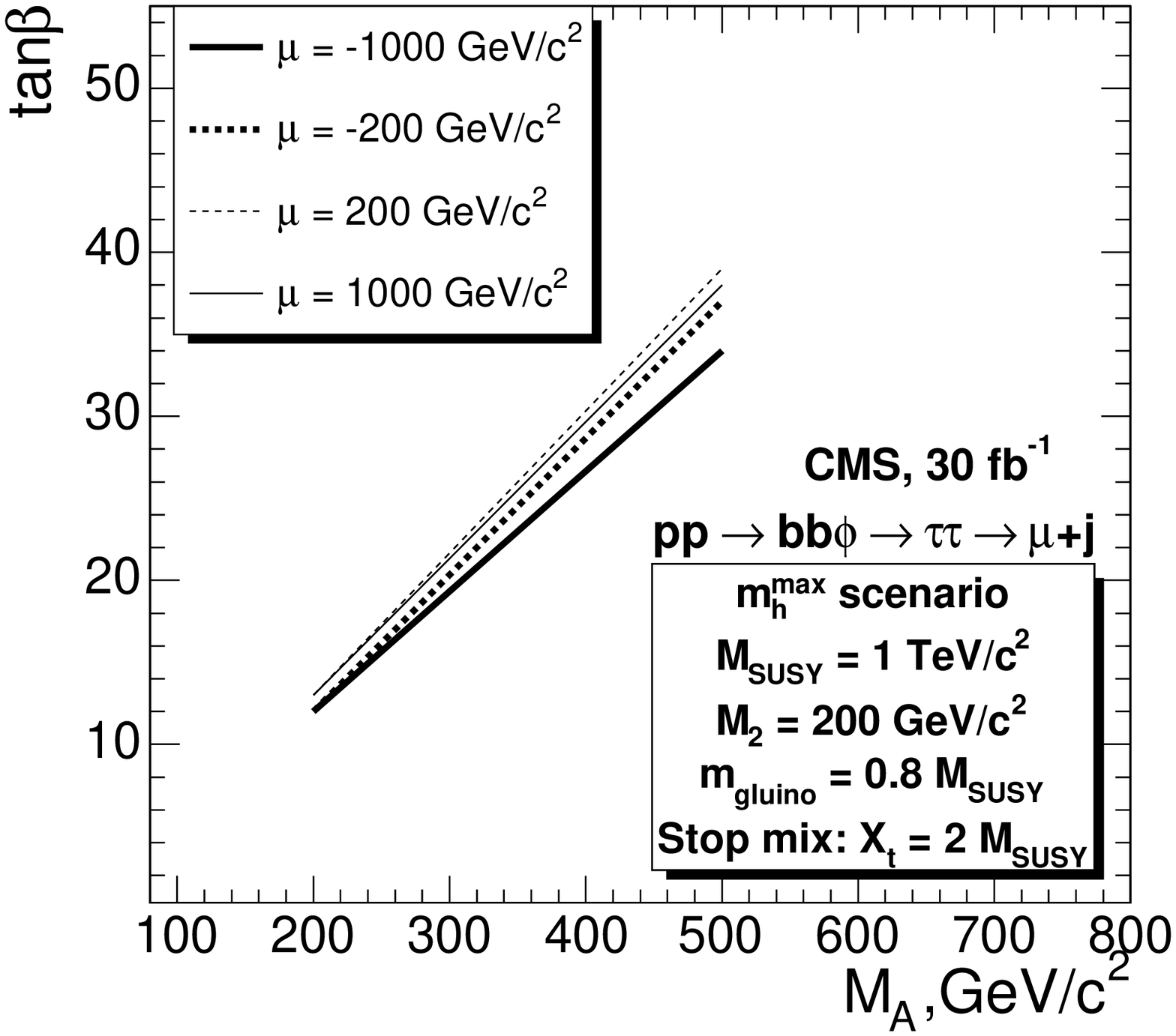}
\EC
\vspace{-1em}
\caption{Variation of the $5 \si$ discovery contours obtained 
in the $\mhmax$ scenario for different values of $\mu$ 
from the channels 
$b\bar b \phi, \phi \to \tau^+\tau^- \to \,\mbox{jets}$ (top),
$\to e + \,\mbox{jet}$ (middle), $\to \mu + \,\mbox{jet}$ (bottom).
}
\label{fig:discovery}
\vspace{-1em}
\end{figure}

\newcommand{\db}{\De_b}
In \reffi{fig:discovery} we show 
the $5 \si$ discovery contours obtained from the process 
$b\bar b \phi, \phi \to \tau^+\tau^-$ for the final states
$\tau^+\tau^- \to \,\mbox{jets}$,  
$\tau^+\tau^- \to e + \,\mbox{jet}$ and 
$\tau^+\tau^- \to \mu + \,\mbox{jet}$.
The 5$\,\si$ discovery contours are
affected by a change in $\mu$ in two ways. Higher-order contributions,
in particular the ones associated with $\db$~\cite{deltamb2}, 
modify the Higgs-boson 
production cross sections and decay branching ratios. 
Furthermore the mass eigenvalues of the charginos and neutralinos vary
with $\mu$, possibly opening up the decay channels of the Higgs bosons
to supersymmetric
particles, which reduces the branching ratio to $\tau$~leptons.

As expected from the discussion of the $\db$ corrections in
\citeres{benchmark3,cmsHiggs}, 
the variation of the 5$\,\si$ discovery contours with $\mu$
can be sizable. In the $\tau^+\tau^- \to \,\mbox{jets}$ channle (top
plot in \reffi{fig:discovery}) a shift up to 
$\De\tb = 12$ can be observed for 
$\MA = 800 \gev$. For low $\MA$ values (corresponding also to lower
$\tb$ values on the discovery contours) the variation stays below
$\De\tb = 3$. In the no-mixing scenario the variation does not exceed 
$\De\tb = 5$. 
The 5$\,\si$ discovery regions are
largest for $\mu = -1000 \gev$ and pushed to highest $\tb$ values for
$\mu = +200 \gev$. 
In the low $\MA$ region our discovery contours are
very similar to those obtained in \citere{benchmark3}. In the high $\MA$
region, $\MA \sim 800 \gev$, corresponding to larger values of
$\tb$ on the discovery contours, our improved evaluation of the 5$\,\si$ 
discovery contours gives rise to a shift towards higher $\tb$ values 
compared to \citere{benchmark3} of about $\De\tb = 8$ (mostly due to
the up-to-date experimental input).

The results for the channel $\tau^+\tau^- \to e + \,\mbox{jet}$ are shown in
the middle plot of \reffi{fig:discovery}. The resulting 
shift in $\tb$ reaches up to $\De\tb = 8$ for $\MA = 500 \gev$.
Finally in the bottom plot of \reffi{fig:discovery} the results for the
channel  $\tau^+\tau^- \to \mu + \,\mbox{jet}$ are depicted. The level
of variation of 
the 5$\,\si$ discovery contours is the same as for the $e + \,\mbox{jet}$
final state.%
\footnote{Since the results of the experimental simulation for this
channel are available only for two $\MA$ values,
the interpolation is a straight line. This may result in a slightly larger
uncertainty of the results compared to the other two channels.}

In \citere{cmsHiggs} it has been shown that the effects visible in
\reffi{fig:discovery} arising from the variation of $\mu$ are a mixture of 
two effects: the change in the bottom Yukawa coupling via $\db$ and the
impact on the heavy Higgs decay channels of possible additional decays
to charginos or neutralinos. The variation of other parameters entering
the radiative corrections is comparably small.


\newpage
\section{Numerical results for the Higgs-boson mass precision}

\begin{figure}[htb!]
\vspace{-0.5em}
\BC
\includegraphics[width=.4\textwidth,height=6cm]{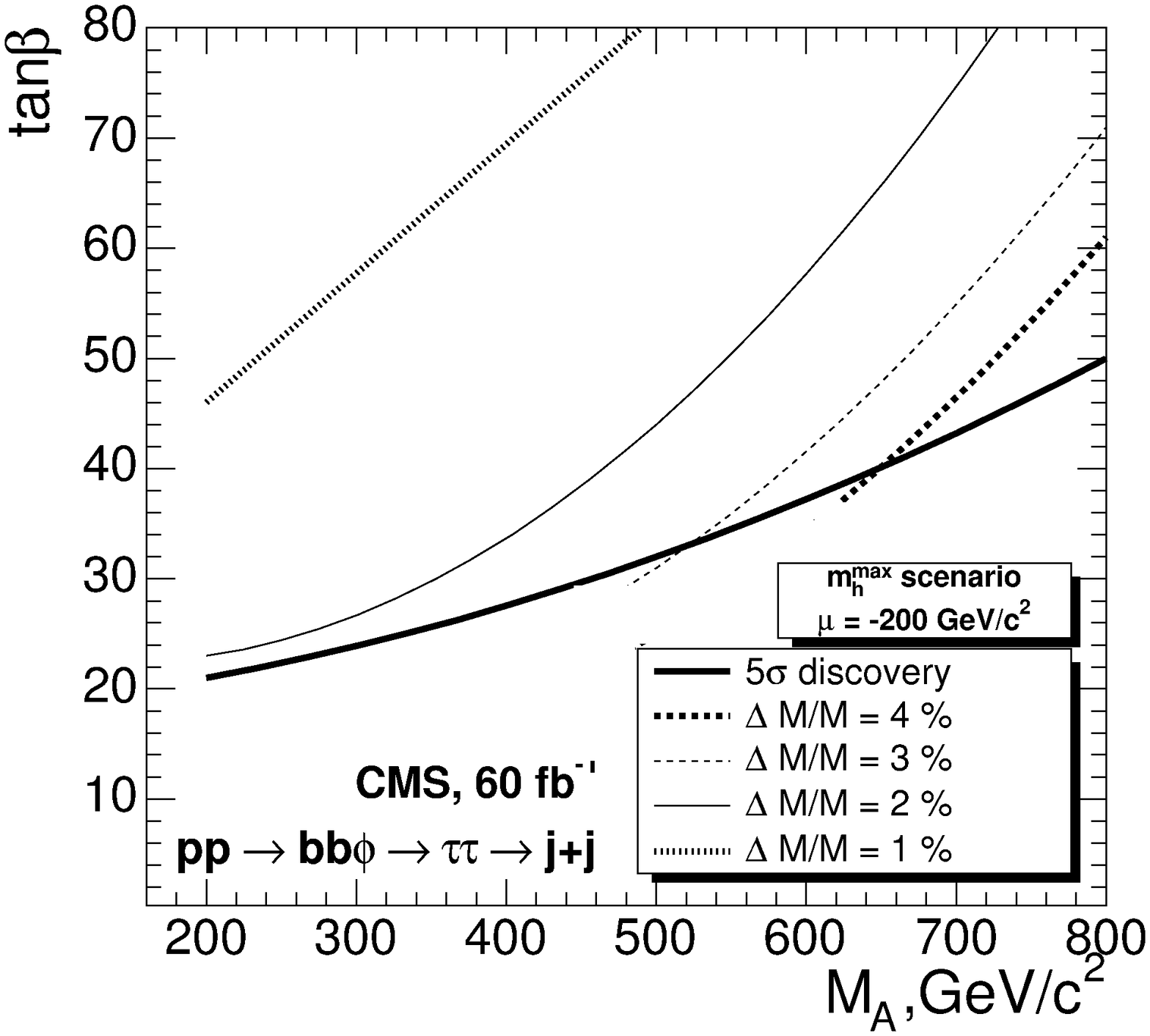}
\includegraphics[width=.4\textwidth,height=6cm]{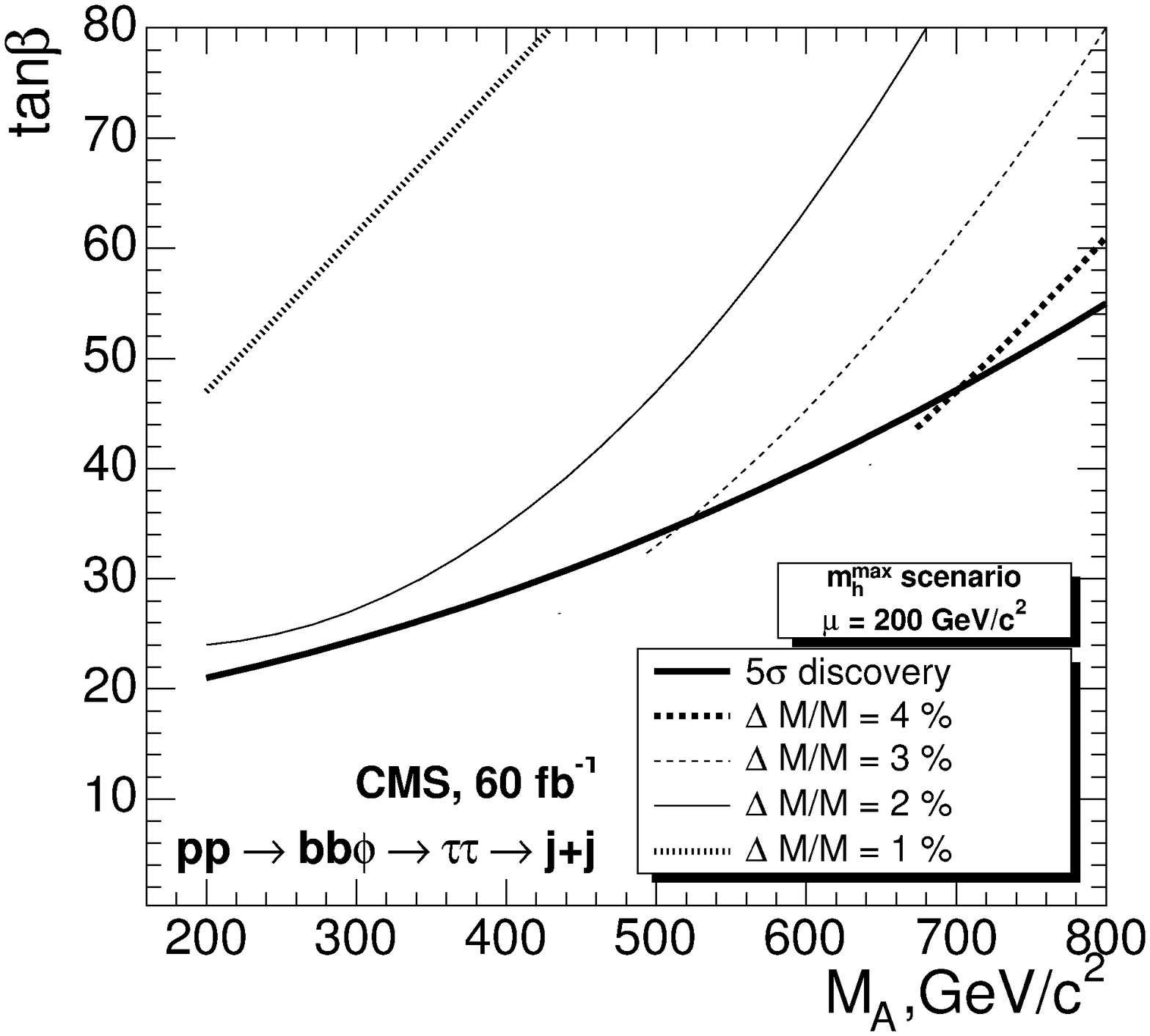}
\EC
\caption{The statistical precision of the Higgs-boson mass measurement
achievable 
from the channel $b\bar b \phi, \phi \to \tau^+\tau^- \to \,\mbox{jets}$
in the $\mhmax$ benchmark scenario for $\mu = -200 \gev$ (top) and 
$\mu = +200 \gev$ (bottom) is shown together with the 5$\,\si$ discovery 
contour.}
\label{fig:jjmass}
\vspace{-2em}
\end{figure}

The expected statistical precision of the heavy Higgs-boson masses is
evaluated according to \refeq{eq:precM}. 
In \reffi{fig:jjmass} we show the expected precision for the mass
measurement achievable from the channel 
$b\bar b \phi, \phi \to \tau^+\tau^-$ using the final state
$\tau^+\tau^- \to \,\mbox{jets}$. Within the 5$\,\si$ discovery 
region we have indicated contour lines corresponding to different values
of the expected precision, $\De M/M$.
The results are shown in the $\mhmax$ benchmark
scenario for $\mu = -200\gev$ (upper plot) and $\mu = +200 \gev$ 
(lower plot).
We find that
experimental precisions of $\De M_\phi/M_\phi$ of 1--4\% are
reachable within the discovery region.
A better precision is reached for larger $\tb$ and smaller
$\MA$ as a consequence of the higher number of signal events in this
region. 
The other channels and other values of $\mu$ discussed above 
yield qualitatively similar results to those
shown in \reffi{fig:jjmass}.

\subsection*{Acknowledgements}

We thank S.~Gennai, A.~Kalinowski, R.~Kinnunen and S.~Lehti for
collaboration on the work presented here.
Work supported in part by the European Community's Marie-Curie Research
Training Network under contract MRTN-CT-2006-035505
`Tools and Precision Calculations for Physics Discoveries at Colliders'.




\end{document}

%% file: susy07HA_abstract.tex
The search for MSSM Higgs bosons will be an important goal at the
LHC. In order to analyze the search reach of the CMS experiment for the
heavy neutral MSSM Higgs bosons, we combine the latest results for the
CMS experimental sensitivities based on full simulation studies with 
state-of-the-art theoretical predictions of MSSM Higgs-boson properties.
The experimental analyses are done assuming an integrated luminosity of 
30 or 60~\ifb.
The results are interpreted as 5$\,\si$ discovery contours in MSSM 
$\MA$--$\tb$ benchmark scenarios. Special emphasis is put on the variation 
of the Higgs mixing parameter $\mu$.
While the variation of $\mu$ can shift the prospective discovery reach (and
correspondingly the ``LHC wedge'' region) by about $\De\tb = 10$, 
the discovery reach is rather stable with respect to the impact of
other supersymmetric parameters.
Within the discovery region we analyze the accuracy with which the 
masses of the heavy neutral Higgs bosons can be determined. 
An accuracy of 1--4\% should be achievable, depending on $\MA$ and $\tb$.